\documentclass[aip,rsi,twocolumn,superscriptaddress,floatfix,10pt]{revtex4-1}

\usepackage{graphicx}
\usepackage{amsfonts}
\usepackage{amssymb}
\usepackage{siunitx}
\usepackage{onlyamsmath}
\usepackage{isotope}

\begin{document}

\affiliation{Physikalisches Institut, Karlsruhe Institute of Technology, Wolfgang-Gaede-Strasse 1, 76131 Karlsruhe, Germany}

\title{A compact ultrahigh vacuum scanning tunneling microscope with dilution refrigeration}
\author{T. Balashov, M. Meyer, W. Wulfhekel}

\begin{abstract}
We have designed and built a scanning tunneling microscope (STM) setup for operation at millikelvin temperatures in ultra high vacuum. A compact cryostat with an integrated dilution refrigerator has been built, that allows measurements at a base temperature of  \SI{25}{\milli\kelvin} in magnetic field up to \SI{7.5}{\tesla} with low mechanical and electronic noise. The cryostat is not larger than conventional helium bath cryostats (23 and 13 liters of nitrogen and helium respectively), so that the setup does not require a large experimental hall and fits easily into a standard lab space.
Mechanical vibrations with running dilution circulation were kept below \SI{300}{\femto\meter\per\sqrt\Hz} by mechanically decoupling the STM from the cryostat and the pumping system.
All electronic input lines were low-pass filtered, reducing the electronic temperature to below \SI{100}{\milli\kelvin}, as deduced from the quasiparticle peaks of superconducting aluminium.
The microscope is optically accessible in the parked position, making sample and tip exchange fast and user-friendly. For measurement the STM is lowered \SI{60}{\mm} down so that the sample ends in the middle of a wet superconducting magnetic coil.
\end{abstract}

\maketitle

\section{Introduction}

Scanning tunneling microscopes (STMs) have made an immense contribution to the development of solid state physics on the nanometer scale \cite{Binnig1982,Bai2000}. In many applications of STM, samples are studied down to the atomic level requiring the operation of the instrument in ultra-high vacuum (UHV). Moreover, cryogenic temperatures are necessary to investigate objects down to single atoms or molecules in order to prevent diffusion \cite{Eigler1990}. STM is capable of measuring the electronic density of state (DOS) at the surface with lateral and energy resolution linking STM observations directly to the quantum mechanics of electronic wave functions \cite{Tersoff1985} or many body effects like in e.g.\ Kondo systems \cite{Madhavan1998} or superconductors \cite{Pan2000}. Additionally, inelastic tunneling processes can be detected in the STM current to obtain atomically resolved information on vibronic excitations e.g.\ in molecules \cite{ITSHo}. STM has been used by many groups to also study spin degrees of freedom, i.e. magnetism, at surfaces and in nanostructures \cite{WiesendangerReview,WulfhekelReview}. To verify spin related origins of observed effects, the application of large magnetic fields is indispensable \cite{Song2010}. As lower temperatures give a higher spectral resolution and a more favorable ratio of thermal smearing versus the Zeeman energy \cite{Assig2013}, there has been a clear trend in this field to go to lower temperatures. Hand in hand with this, the experiment setups became not only bulkier but also exceedingly expensive, especially when entering the domain of dilution refrigeration.  Only a small number of comparable systems has been mentioned in the literature (see Table~\ref{table}) and all of them require large experimental halls, consume large amounts of liquid helium and/or have no optical access to the tunneling junction, which is necessary e.g.\ to access small superconducting samples with lateral sizes about \SI{1}{\milli\meter}. They mostly require relatively long cooldown time and have complex mechanical transfer systems to exchange samples and tips. 

Here we present our efforts to bring efforts associated with an STM operated at \SI{25}{\milli\kelvin} back down to the scale of conventional setups operated with a helium bath cryostat. The setup presented here combines a compact design with internal damping of the microscope with an easy and quick sample transfer without compromising on the essential performance marks like the base and the electronic temperature, helium consumption and standing times as well as vibration level.   
 
\begin{table*}[hb]
\begin{tabular}{cccccccc}


Location & U. of Tokyo~\cite{Kambara2006} & NIST~\cite{Song2010} & MPI Stuttgart~\cite{Assig2013} & Princeton U.~\cite{Misra2013} & U. of Maryland~\cite{Roychowdhury2014} & Radboud U.~\cite{Allwoerden2018} & KIT \\
Publication year & 2006 & 2010 & 2013 & 2013 & 2014 & 2018 & 2018\\
Temperature & \SI{30}{\milli\kelvin} & \SI{10}{\milli\kelvin} & \SI{12}{\milli\kelvin} & \SI{20}{\milli\kelvin} & \SI{30}{\milli\kelvin} & \SI{30}{\milli\kelvin} & \SI{25}{\milli\kelvin}\\
Electronic temperature & $>$ \SI{350}{\milli\kelvin} &  & \SI{38}{\milli\kelvin} & \SI{250}{\milli\kelvin} & \SI{184}{\milli\kelvin} & \SI{195}{\milli\kelvin} & \SI{95}{\milli\kelvin}\\
Axial magnetic field & \SI{6}{\tesla} & \SI{15}{\tesla} & \SI{14}{\tesla} & \SI{14}{\tesla} & \SI{14}{\tesla} & \SI{9}{\tesla} & \SI{7.5}{\tesla}\\
LHe tank volume & \SI{74}{\liter} & \SI{250}{\liter} & \SI{160}{\liter} & \SI{72}{\liter} & \SI{140}{\liter} & \SI{80}{\liter} & \SI{12}{\liter} \\
LHe standing time & \SI{3}{\day} & \SI{11}{\day} &  & \SI{4}{\day} &  & \SI{>4}{\day} & \SI{6}{\day}\\
Visible junction & no & no & no & no & no & no & yes\\
Turnaround time & \SI{3}{\hour} &  & \SI{15}{\hour} & \SI{>6}{\hour} & \SI{15}{\hour} &  & \SI{2}{\hour}\\
Coarse motion &  & \SI{3}{\mm} XY & no & \SI{7}{\mm} X & no & no & \SI{5}{\mm} X \\
\end{tabular}
\caption{UHV dilution refrigerated STM setups with sample exchange found in literature and their key parameters. No entry indicates that the value is not given in the referenced paper.} \label{table}
\end{table*}

\section{UHV system}

\begin{figure*}[bh]
\includegraphics[width=0.8\textwidth]{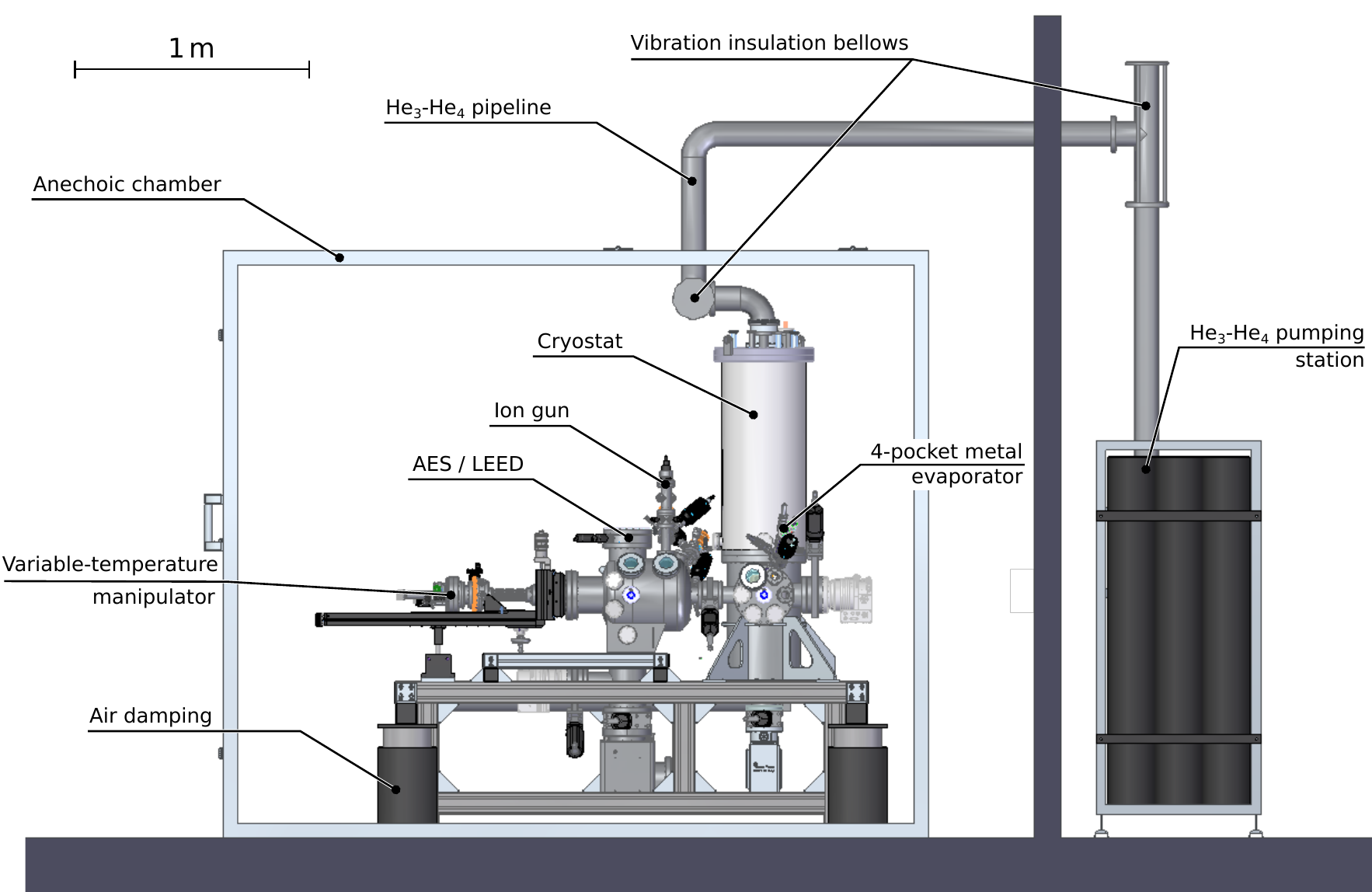}
\caption{Schematic drawing of the experimental setup including the UHV system with cryostat and  STM, and an anechoic chamber in the laboratory and the pumping system in an adjacent service room.}\label{whole-setup}
\end{figure*}

The setup is split into three vacuum chambers, a primary one for the cryostat and the STM itself, a second one for in situ sample preparation and surface analysis and a load-lock for sample and tip transfer from ambient conditions to the preparation chamber (see Fig.~\ref{whole-setup}). All three chambers are equipped with individual turbo molecular pumps so that they can be vented and pumped down to vacuum again independently for individual service, if necessary.
Both the STM and the preparation chamber are equipped with ion getter pumps to achieve base pressure of \SI{2e-10}{\milli\bar}. The preparation chamber additionally contains a titanium sublimation pump and the STM chamber a home-built \SI{300}{\liter\per\second} pump based on non-evaporable getter stripes from SAES \footnote{SAES group: www.saesgetter.com}. During STM operation, all three turbo molecular pumps need to be switched off.

The load-lock additionally allows the connection of a home-built vacuum suitcase for investigating samples prepared elsewhere without breaking the vacuum \cite{Jandke2018}. It can also house a Knudsen cell for sublimation of organic molecules and a quartz micro balance. 

The preparation chamber houses a sputter gun for ion etching of sample surfaces and an electron beam heater to anneal samples up to \SI{1800}{\kelvin}. The surface crystal structure can be investigated with the installed low energy electron diffraction (LEED) optics. The same optics can be used to record Auger electron spectra of the surface for chemical analysis.  

Several electron beam evaporators allow computer controlled and simultaneous deposition of up to four materials onto the sample \footnote{SPECS GmbH: www.specs.de}. The samples are handled via the main manipulator allowing lateral movement between the preparation and STM chamber, but also rotation to face the sample surface towards the different instruments. It is equipped with a flow cryostat that allows to cool the sample controllably down to \SI{90}{\kelvin} using liquid nitrogen or \SI{20}{\kelvin} using liquid helium \footnote{VAb Vakuum--Anlagenbau GmbH: www.vab-vakuum.com}.

The load-lock, preparation chamber and STM chamber all have sample racks to store between 3 and 9 samples. This not only allows to work on different samples but also to have service times for each chamber individually without exposing samples to air. Two wobble sticks in the preparation and the STM chamber are used to transfer samples and tips between the manipulator, the load lock, the racks and the STM.

The complete setup is contained in an anechoic room to minimize the influence of external sounds on the measurement. The rotary pump and the mixture pumping system are installed in a separate room outside the main lab (see Fig.~\ref{whole-setup}).

\section{Design of the cryostat}

\begin{figure}
\includegraphics[width=1\columnwidth]{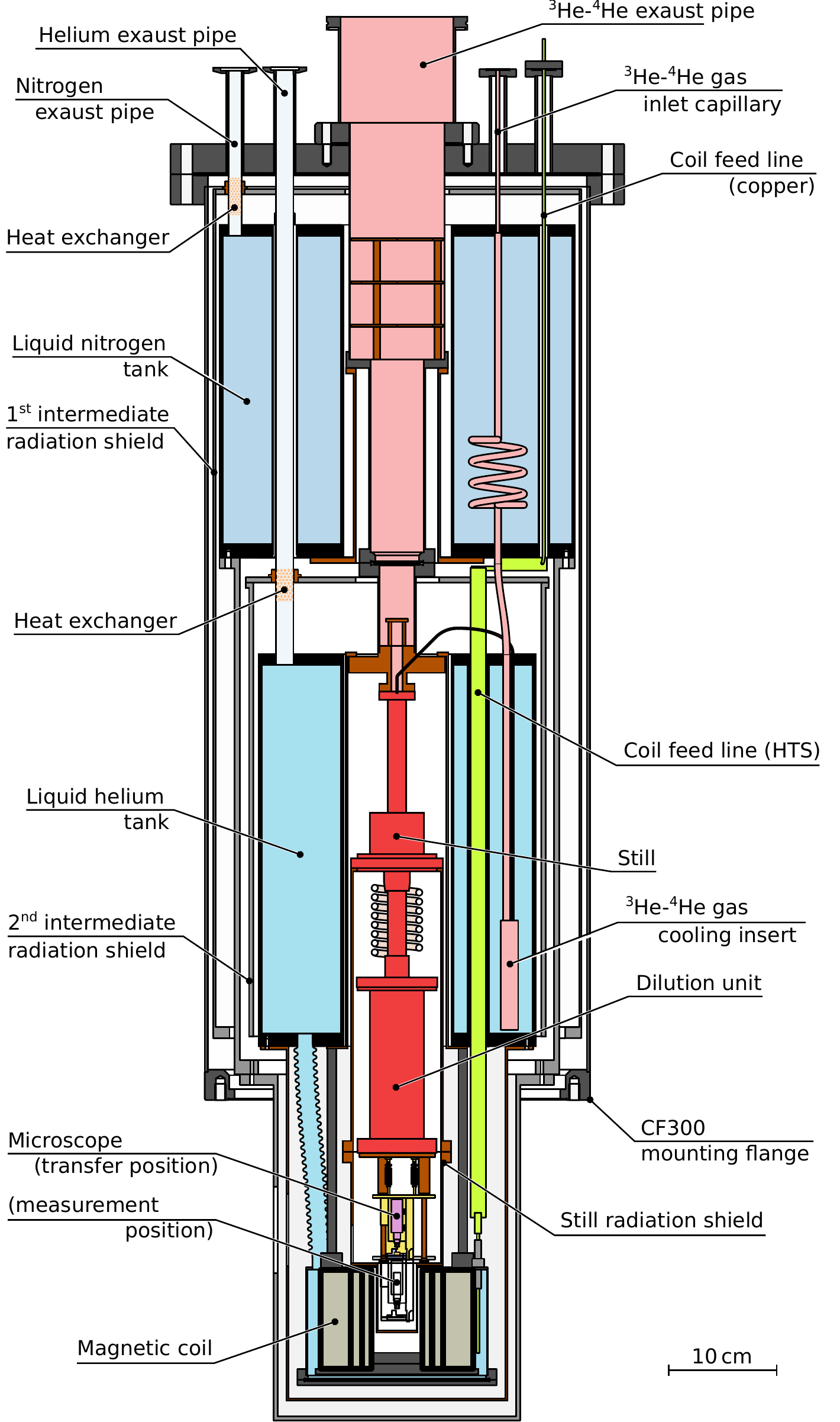}
\caption{Simplified sketch of the cryostat containing the liquid nitrogen and helium tanks (blue), the dilution insert (red), the gas circulation system (pink), the STM and the superconducting magnet. The current supply lines are colored green.}\label{fig_Cryo}
\end{figure}

For compatibility with other vacuum system in our lab, the cryostat is built on a standard CF300 flange and measures \SI{85}{\centi\meter} in height from the bottom to the top flange, with additional \SI{30}{\centi\meter} from the bottom flange to the bottom of the radiation shield (the cryostat can be exchanged with the Joule-Thomson cryostat of our second setup \cite{Zhang2011}). The whole setup thus measures mere \SI{3}{\meter} from the floor to the top including the pipes used to transfer \isotope[3]{He}-\isotope[4]{He} mixture in and out of the cryostat. 

To achieve these small dimensions in a cryostat for UHV, several design rules needed to be implemented. Most importantly, we needed to deviate from the typical design of a dilution insert, which is inserted into a conventional, super-insulated bath cryostat. Instead, liquid nitrogen (\SI{23}{\liter}) and helium (\SI{13}{\liter}) tanks are stacked on top of each other directly in UHV. Each of them is insulated with radiation shields extending around the inner regions as shown in Fig.~\ref{fig_Cryo}. 
Every tank is mounted on three pipes, with one used for filling, one as exhaust and a reserve one. The exhaust pipes both contain an open pore copper foam heat exchanger, so that the escaping cold gas is used to cool the intermediate radiation shield to minimize heat input. The two intermediate radiation shields stabilize at 180 and \SI{28}{\kelvin}. Note that these temperatures are significantly lower than those expected from thermally floating or passive shields that would stabilize between the warm and cold side by the equilibrium radiation flux. This demonstrates the high efficiency of the heat exchangers. Additionally, the liquid nitrogen tank is filled with fine aluminium wool to prevent bubbling and by this vibrations. Note that all pipes are attached to the actively cooled shields in order to minimize boil off due to heat conduction. Similarly, the \SI{100}{\milli\meter} pipe in the middle of the cryostat, which serves as part of the \isotope[3]{He} pumping line, is stepwise cooled by the two baths and two active radiation shields. 
This way, the total heat input to the nitrogen tank can be kept low and was estimated to be \SI{13}{\watt}, with half from radiation and half from thermal conduction.

Most of the cryostat is bakeable above \SI{150}{\degreeCelsius} with the exception of the dilution insert and the magnetic coil, that are kept slightly below \SI{80}{\degreeCelsius} during the bakeout.

\subsection{Dilution refrigerator}

The dilution refrigerator is a commercially available insert from BlueFors Cryogenics~\footnote{BlueFors Cryogencis: www.bluefors.com} with only minor construction changes to fit our system. The insert is mounted in a \SI{100}{\milli\meter} opening inside the LHe tank. The \isotope[3]{He}-\isotope[4]{He} gas mixture is supplied through a \SI{1}{\milli\meter\squared} capillary, precooled to \SI{77}{\kelvin} in the liquid nitrogen tank, then to about \SI{28}{\kelvin} on the helium radiation shield and finally to \SI{4.2}{\kelvin} in the liquid helium tank. It is then further cooled via a third open pore copper heat exchanger on the return line of the dilution circulation, so that the temperature of the gas reaching the dilution insert is normally below \SI{4}{\kelvin}. In the dilution unit, first a Joule-Thomson expansion stage further cools down the gas to condense it. No \SI{1}{\kelvin}-pot is needed this way, simplifying the design of the cryostat and avoiding vibrations of the superfluid helium. The incoming mixture is then heat exchanged with the outgoing one via a continuous and then a series of discrete heat exchangers. Routinely, temperatures below \SI{10}{\milli\kelvin} can be reached with this mixing unit at the mixing chamber. Test of the cryostat without the STM installed revealed a base temperature of \SI{8}{\milli\kelvin}.

The returning gas is pumped out through the central pipe. The pumping station is installed outside the main laboratory. To reduce the height of the complete setup, the temperature gradient is not only in the vertical or radial direction, but the temperatures of the liquid nitrogen and the liquid helium bath are brought upwards to the central dilution part. For example, the \SI{77}{\kelvin} point of the central pumping line is brought upwards to roughly the middle of the nitrogen tank by a thick copper pipe supporting the inner pumping line and its radiation baffle. The \SI{4.2}{\kelvin} point is even brought up to the top of the He tank in a similar manner.  

\subsection{Magnetic coil}

A wet superconducting \SI{7.5}{\tesla} magnetic coil is mounted below the LHe tank and is connected to it by a pipe. 
The advantage of the design of a separate helium tank for the magnet is that the STM and its sample rack are visible to the user, when the STM itself is in transfer position. The STM can then be lowered into the magnet for operation (see below). The design further allows to easily exchange the magnet, e.g.\ exchange a solenoid for a vector magnet, without the need to reconstruct the cryostat.

\begin{figure}
\includegraphics[width=\columnwidth]{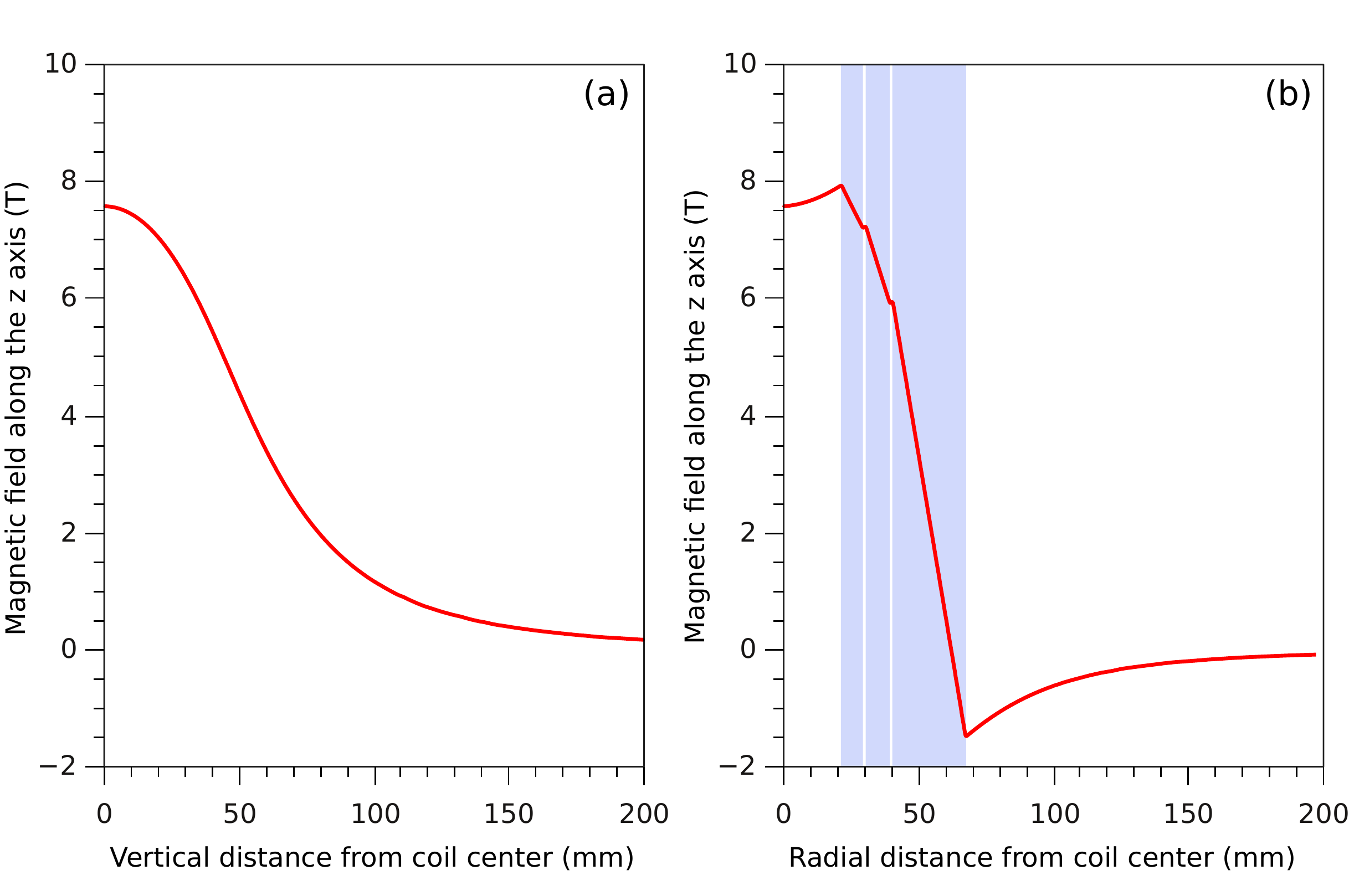}
\caption{Calculated magnetic field distribution inside the coil in axial (a) and radial (b) directions. Blue rectangles in the right plot indicate the width and location of the three coil sections with different wire diameter.} \label{magnet}
\end{figure}

The NbTi-based coil was built in-house using wire from Supercon Inc.~\footnote{Supercon Inc.: www.supercon-wire.com} and optimized for maximal magnetic field in the available geometry in combination with a current below \SI{35}{\ampere}. To satisfy these limits, we have split the coil into three sections of different wire diameter. The design was optimized using an analytical model that was then crosschecked with finite element simulations \cite{MaerklPhD}. The result is shown in Figure~\ref{magnet}. The outer coil with the thinnest wire creates most of the field up to about 50\% of its critical current at the achieved fields \cite{SuperconductingMagnetsWilson}. The inner two coils with thicker wires can sustain their superconducting state up to higher fields, which is used to bring the field up to about \SI{7.5}{\tesla}. The glue used to make the coils has a glass temperature of about \SI{100}{\degreeCelsius} limiting the baking temperature of the magnet.

To minimize cryogent losses for cooling the current line, it is split into two parts. A first part bring the current from room temperature to liquid nitrogen using shielded AWG~11 copper wire. A second part brings the current from the liquid nitrogen tank to the coil at \SI{4.2}{\kelvin} via superconducting 2G HTS band wires out of rare earth barium copper oxide~\footnote{SuperPower Inc.: www.superpower-inc.com} with a critical temperature above \SI{80}{\kelvin} and a critical current at \SI{77}{\kelvin} of \SI{280}{\ampere}. These high-temperature superconductors on a thin steel tape have negligible thermal conduction. Note that in case this superconductor quenches, the energy stored in the magnetic coil will be sufficient to evaporate the thin superconducting coating. Thus, special care has to be taken to avoid exposure of the tape to thermal radiation and the temperature of the contact between normal and superconducting parts of the current line is monitored by a separate Pt1000 thermometer.

\subsection{The microscope}

\begin{figure}
\includegraphics[width=0.9\columnwidth]{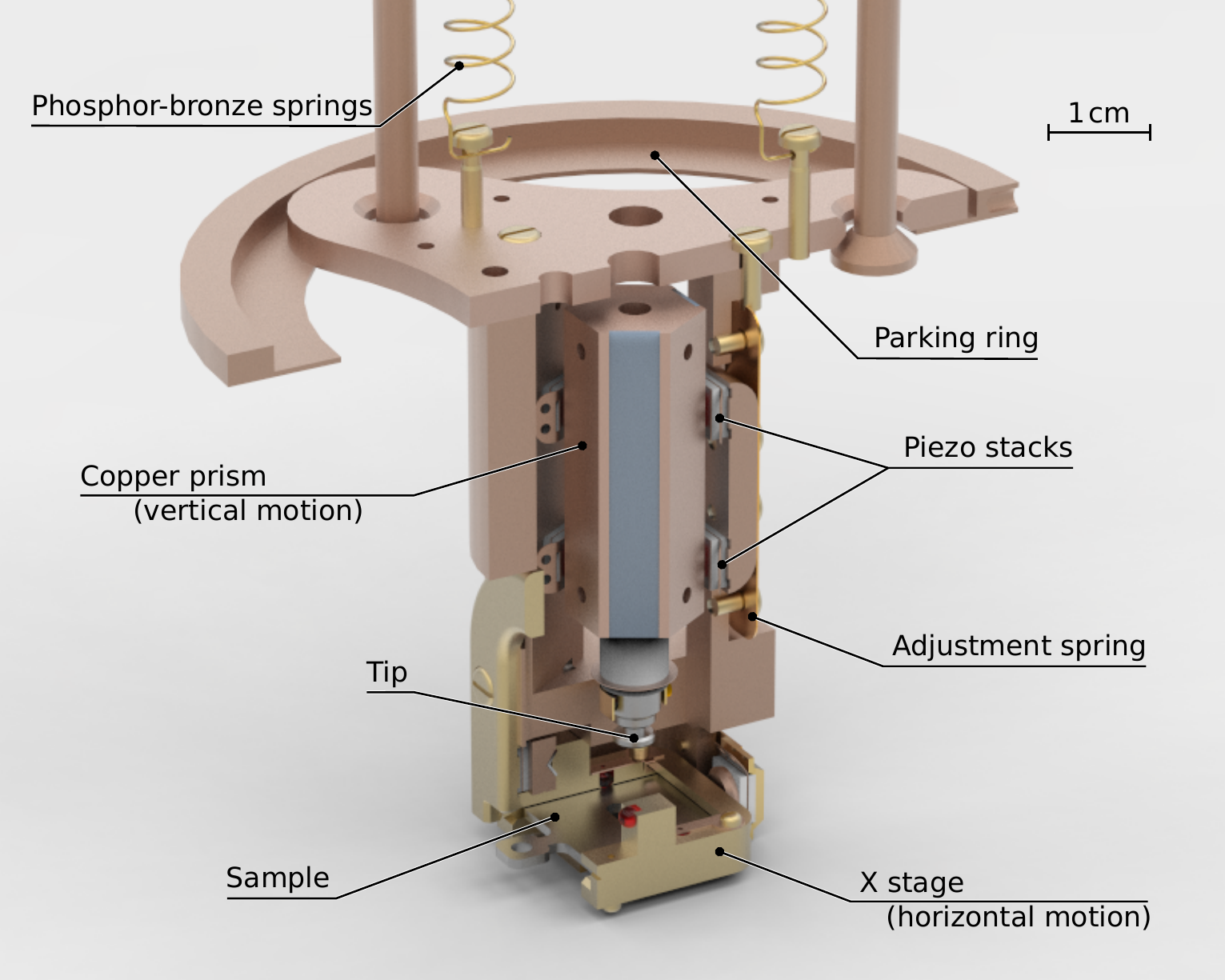}
\caption{3d render (cut) of the STM and parts of the suspension. The copper ring serves to lift the STM up to the parked position and does not touch the STM in the measurement position. The STM body has a diameter of \SI{30}{\mm}.} \label{STM}
\end{figure}

The design of the STM was guided by two requirements. First, in situ exchange of the sample and the tip was desired with optical control of the junction and the ability to deposit materials on the sample at temperatures below \SI{10}{\kelvin}. Second, the cross-section was to be kept small to minimize the bore diameter of the magnetic coil (\SI{44}{\mm}) and maximize the magnetic field for the given coil dimensions. The resulting microscope uses $15\times\SI{15}{\milli\meter\squared}$ sample plates compatible with those used in our previous design \cite{Zhang2011} and allows translational motion in $x$ direction within a \SI{5}{\milli\meter} range.  This motion is also used to exchange tips between the STM scanner and transfer plates for STM tips. The STM scaffolding is mounted directly on the bottom of the mixing chamber and features two stable positions for the STM. In the transfer position the STM is thermally connected to the still volume and is physically and optically accessible through the sliding doors in the liquid nitrogen and liquid helium shields. These can easily be opened with the same wobblestick that is used to put in or take out samples from the STM.  While rotating the still door with the wobble stick, the still door is closed and the STM is lowered at the same time. In the working position the STM is located in the center of the coil such that the sample is situated in the region of maximal field. In this position the STM is thermally connected only to the \SI{10}{\milli\kelvin} pot by thin, high purity silver wires and reaches temperatures down to \SI{19}{\milli\kelvin}, as controlled by a calibrated RuO$_2$ thermometer mounted on the STM body. A \SI{10}{\kilo\ohm} resistor acting as a heater is mounted on the STM body and can be used to controllably increase the STM temperature up to \SI{1}{\kelvin}.

Most of the STM parts are made of copper and gold-plated to ensure fast and homogeneous cooling (see Fig.~\ref{STM}). Small parts where extensive wear or tension is expected, such as the sample stage and all screws, were made of beryllium-copper alloy. STM cabling inside the cryostat is made out of shielded stainless steel cables pre-cooled at \SI{77}{\kelvin} and \SI{4}{\kelvin} plus additionally at the still shield by induction-less winding on copper cylinders. To make exchange of the STM head possible, the cabling is broken at the base plate with a vacuum-compatible connector. Inside the STM head stainless steel cables are used for tunneling current and bias voltage, while a \SI{0.3}{\mm} flat flexible polyimide cable~\footnote{Molex LLC: www.molex.com} is used for piezo connections.
Coarse motion in $z$ and $x$ directions was realized with slip-stick motors on polished sapphire surfaces. The gliders on these surfaces are polished alumina ceramics. To ensure low temperatures also in the moving parts of the STM, the central $z$-prism holding the scanner piezo was also made from copper with the gliding surfaces of sapphire glued in. The prism is thermally connected to the body of the STM by a flexible braided copper wire. Similarly, to ensure low temperatures at the tip and electrical insulation from the scanner piezo, the insulation between both is performed by two insulating rings with a copper ring in between. The copper ring is thermally anchored to the prism.

\subsection{Noise considerations}

Several mechanical noise sources are of importance for a low-temperature STM setup. First, mechanical excitations from outside the system need to be considered. For our system this includes acoustic waves, that are damped by the anechoic chamber, the vibrations of the building, that are damped by the passive laminar flow insulation legs and the vibrations caused by the pumps of the \isotope[3]{He} circulation system, that normally need to be running during the measurement. The last problem is addressed in several steps. On one hand, the pipe connecting the the pumping system to the cryostat has two double-bellow decouplings. On the other hand, the pumps themselves are mounted on shock-absorbing rubber posts, minimizing vibration transmission to the pipe in the first place. Additionally, a Helmholtz-resonator is mounted on the pumping line to selectively damp the frequency of the mechanical pumps.

Second, internal resonances play a role. Most important ones are the bending modes of the two cryogenic tanks and the radiation shields as well as the internal modes of the STM body. The former were damped in vacuum by additional stiff supports on the bottom end of the bell shaped shields, while the latter has been confirmed during the design stage to lie above \SI{5}{\kilo\hertz}. To further minimize the vibration transmission to the STM, it hangs on three soft springs and is additionally damped with UHV compatible Kapton foam when hanging in the lower, operation position. In this way, a separate foundation for the lab could be avoided.

Electronic noise sources include electromagnetic waves coupling to the wiring of the instruments, ground loops and internal noise of the bias voltage supply and the tunneling current amplifier. Additionally, at mK temperatures, radiation of thermal microwave photons down the electric lines into the tunneling junction matter.

To avoid coupling to electromagnetic waves, the anechoic chamber also serves as a Faraday cage and all electrical signals enter or exit the box in a shielded way. To avoid ground loops, one needs to focus on the tunneling bias and tunneling current lines. They were first of all routed in parallel inside the STM. Secondly, the commercial transimpedance amplifier from Femto \footnote{FEMTO Messtechnik GmbH: www.femto.de} measures the current versus ground of the input channel. The latter is connected by an internal resistor to the ground of the output channel, which itself is connected to the machine ground. Thus, the tunneling current flows from the sample bias line via the amplifier to the ground of the instrument. The output enters the STM electronics, however, on a floating input. The only important ground loop that needs to be considered is that between the STM electronics and the machine ground via the shielded sample bias cable and via all other connections. It is thus easiest to cut the path of the bias voltage to remove all loops acting on the tunneling current. This is done by inserting an instrumental amplifier between the STM electronics providing the sample bias, and the cryostat close to the electrical feedthrough of the tunneling current. We used a home-built instrumental amplifier powered by a separate battery. Further, the instrumental amplifier has a second order low pass filter with a cutoff frequency (\SI{-6}{\decibel}) of about \SI{30}{\kilo\hertz}. It further allows to passively divide the bias voltage directly at the exit of the circuit by 10 or 100. This allows to bring the voltage noise down to \SI{3}{\nano\volt\per\sqrt\hertz}.

All electrical signals are filtered by high frequency $\pi$-filters from Spectrum Control \footnote{Spectrum Control Inc.: www.SpecEMC.com} with -7 and \SI{-60}{\decibel} at 1 and \SI{30}{\mega\hertz}, respectively. Only the tunneling current line cannot be filtered as its capacity needs to be as low as possible to avoid the input voltage noise of the transpedance amplifier to be converted into current noise. Coarse motion and heaters were connected with simple steel cables while fine motion, current and bias were connected with UHV compatible and shielded steel cables \footnote{Cooner Wire Inc.: www.coonerwire.com} of a low capacitance of only 39 pF per meter. The cables have been analyzed with a vector network analyzer showing a damping of 40 and 90 dB per meter at 2 and 8 GHz, respectively. Thus in their way down to the still, the cables also act as filters to lower the electronic temperature. Finally, the bias voltage is filtered at still temperature by a UHV compatible powder filter made by A. Lukashenko \cite{Lukashenko2008}.  

\section{Experiments}

\subsection{Cryostat performance}

\begin{figure}
\includegraphics[width=0.8\columnwidth]{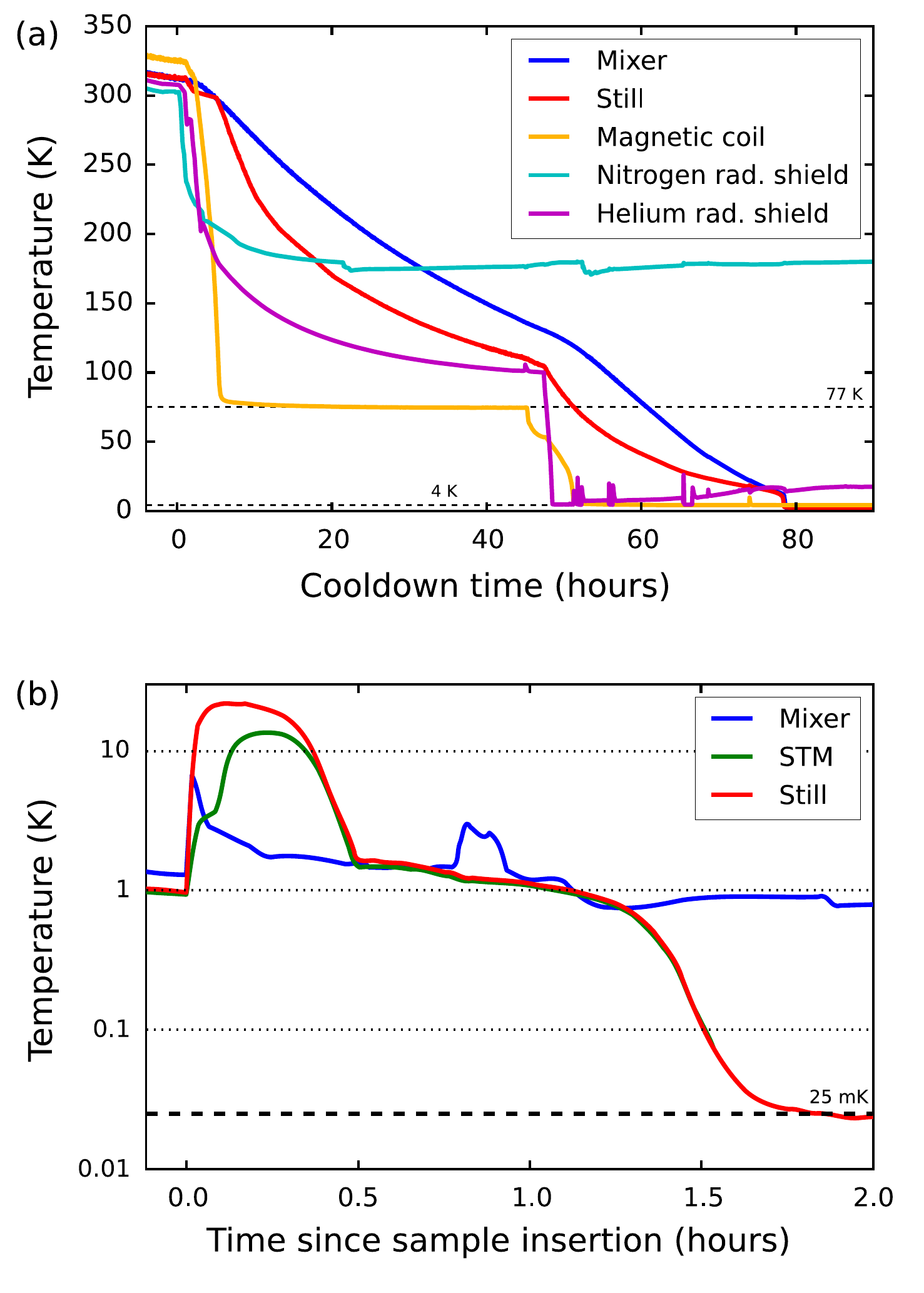}
\caption{(a) Temperatures of critical parts of the setup during cooldown to \SI{1}{\kelvin} after baking. (b) STM cooldown after insertion of a new sample. Due to evaporation of \isotope[3]{He}-\isotope[4]{He} mixture, most of the cooldown time is spent to re-condense it.} \label{cooldown}
\end{figure}

\paragraph{Cooling down}
After exposure of the instrument to ambient conditions, the STM chamber is evacuated and baked for about \SI{48}{\hour} until the pressure is in the range of \SI{1e-7}{\milli\bar} followed by cooling to room temperature with a flow of compressed air though the different tanks. A total time of \SI{80}{\hour} is required to cool down to the base temperature. A trace of the temperatures of several points during the cooldown process is presented in Fig.~\ref{cooldown}a. First, both the nitrogen and the helium tank are filled with liquid nitrogen and rapidly cool down the magnet. The outgoing nitrogen gas cools both the nitrogen and helium radiation shields. After about \SI{20}{\hour} the final temperature of the nitrogen radiation shield of \SI{\approx180}{\kelvin} is reached. The still is coupled by a passive gas heat switch to the helium tank cooling down the still and the mixing chamber. After about \SI{48}{\hour}, the liquid nitrogen is blown out from the helium tank and liquid helium is inserted. Immediately, the helium radiation shield cools down to nearly \SI{4.2}{\kelvin} by the cold exhaust helium gas and the magnet follows quickly after. At about \SI{20}{\kelvin} the passive gas switch of the still transits to the off state, the helium evaporation rate goes down and the helium radiation shield starts to warm up to its equilibrium temperature. Once the still cools down below \SI{13}{\kelvin}, the circulation of the mixer can be started and the Joule-Thomson cooling quickly reduces the still temperature below the condensation temperature of the mixture. It takes about \SI{3}{\hour} to condense all the mixture and start the dilution cooling. 

\paragraph{Cryogent standing time} The standing time of the system depends on the regime in which it is operated. If the dilution unit is not in use such that no circulation takes place, the standing time of the nitrogen tank is larger than \SI{64}{\hour} and that of helium \SI{150}{\hour}.  If, however, there is mixture circulating, the load on the tanks increases, depending on the gas flow. While the increase in nitrogen consumption is minute, condensing the mixture costs about one third of the capacity of the helium tank.
Once the mixture has condensed and the cryostat has reached a steady state at the lowest temperature, the helium consumption drops again resulting in a standing time of \SI{59}{\hour}. In this state, the capacity of the nitrogen and helium tank are balanced, maximizing the measurement times of large spectroscopic maps for the designed size of the cryostat.

\paragraph{Sample exchange} The dilution unit was designed to tolerate large heat input on the mixing chamber without danger.
Thus, the still shield door can be opened with the STM at base temperature, and samples (and STM tips) can be removed or inserted into the STM.
To minimize the heat load they are pre-cooled to \SI{\approx90}{\kelvin} on a sample rack on the nitrogen shield before insertion into the STM. The transfer procedure will then shortly warm up the microscope to \SI{30}{\kelvin}, and evaporate a fraction of the mixture. Fully recondensing the mixture takes between one and two hours, after which the STM can be cooled down to base temperature within an hour. A typical time trace of the temperatures of the STM, the mixing chamber and the still is shown in Fig.~\ref{cooldown}b, in which the experiment reaches \SI{1}{\kelvin} after \SI{30}{\minute} and its base temperature after only \SI{90}{\minute}. As sample exchange is done under direct vision with a  wobblestick, it is fast and easy and sample turnaround time is not significantly larger than in \SI{4}{\kelvin} STM designs. This eliminates one of the important constrains in using STM at mK temperatures.

\subsection{Mechanical and electronic noise}

\begin{figure}
\includegraphics[width=0.9\columnwidth]{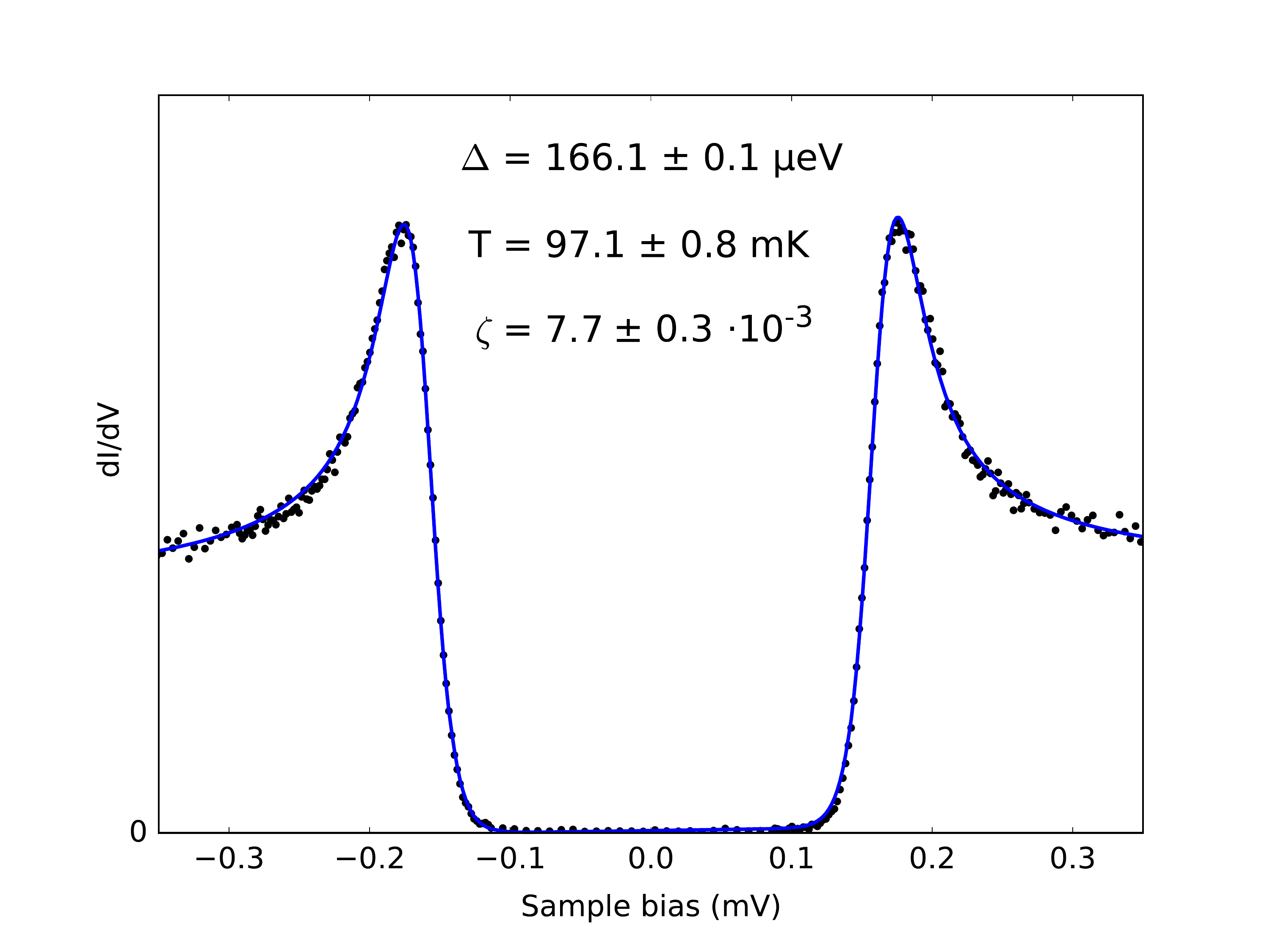}
\caption{Differential conductance of Al(111) at \SI{29}{\milli\kelvin}, measured with a normal conducting W tip, fitted with the quasiparticle DOS in the Maki formalism (see ref.~\cite{Assig2013, Maki1964}.)}\label{fig_Al}
\end{figure}

To analyze the energy resolution of the setup we follow M.~Assig et al.~\cite{Assig2013} in measuring and fitting the density of states of superconducting Al(111). In this experiment, we used a W tip, which should be in the normal state at the STM temperature and an atomically clean surface of a bulk Al(111) single crystal. To not spoil the resolution, the tunneling current was measured directly, without a lock-in amplifier and modulation of the sample bias, and the density of states was obtained by numerically differentiating the $I(V)$ curve. The resulting DOS (see Fig.~\ref{fig_Al}) is fitted with the BCS density of states including spin-orbit effects, i.e.\ with the Maki equation~\cite{Assig2013, Maki1964}. This neglects the coupling to photons via the P(E) theory (for details see Ref.~\onlinecite{Ast2016}) and assumes a superconducting gap $\Delta$ that is isotropic on all three bands cutting the Fermi energy \cite{Blackford1976}. Thus, this is an upper limit of the electronic temperature of the system. The fitted effective temperature is \SI{97.1\pm0.8}{\milli\kelvin}, corresponding to an energy resolution of \SI{\approx30}{\micro\electronvolt}.

\begin{figure}
\includegraphics[width=0.9\columnwidth]{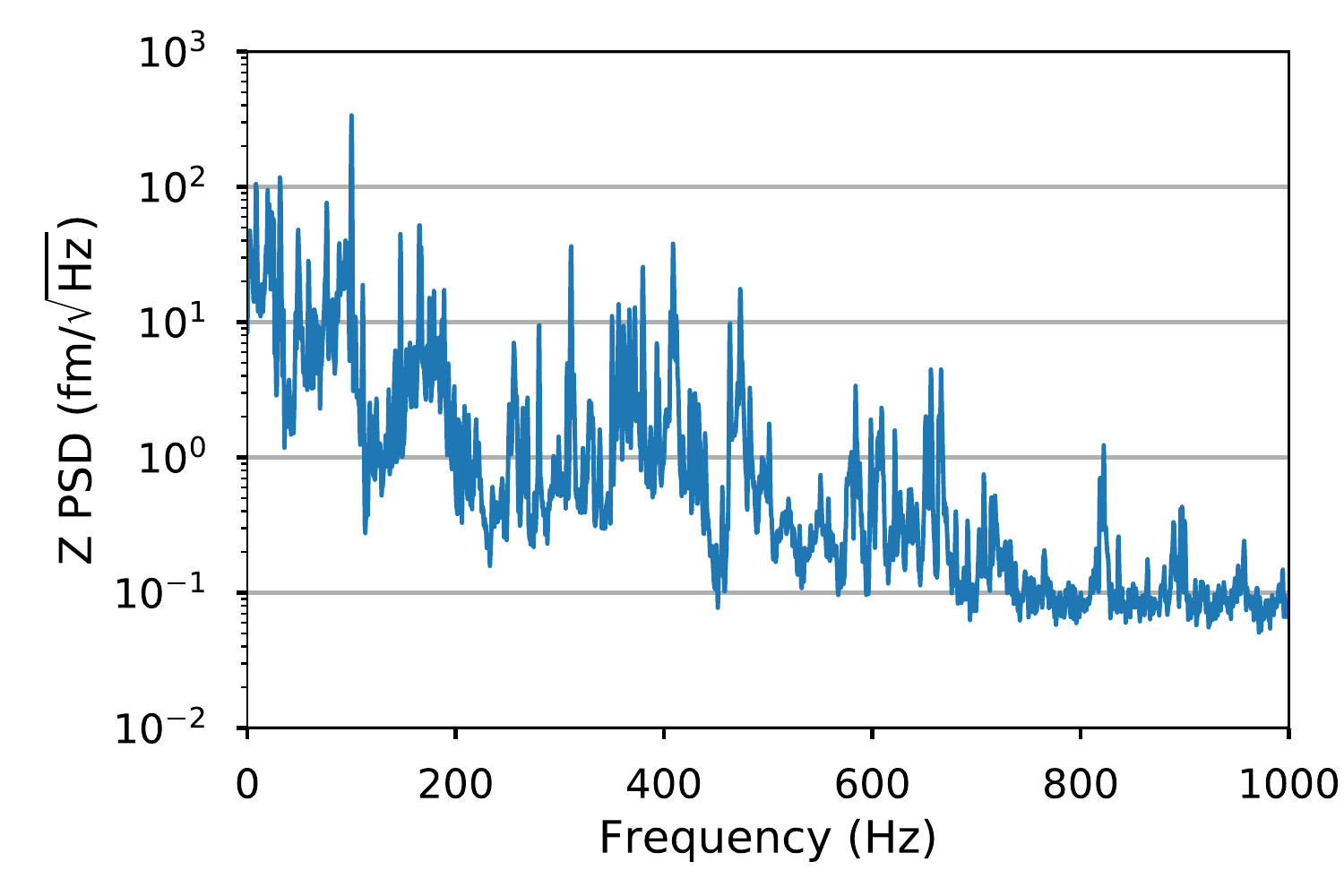}
\caption{Power spectrum of Z noise in tunneling conditions with feedback on at \SI{29}{\milli\kelvin} recorded with a W tip on Al(111).}\label{vibration}\label{fig_noise}
\end{figure}

On the same surface, the vibrational noise of the system was tested by recording time traces of the $z$ position in feedback. The cut-off frequency of the feedback loop was set to \SI{1}{\kilo\hertz} during this measurement. Figure~\ref{vibration} shows the power spectrum of this trace as function of frequency. Besides a sharp peak at \SI{100}{\hertz} of slightly above \SI{300}{\femto\meter\per\sqrt\hertz}, which also contains electronic noise, all peaks are below \SI{120}{\femto\meter\per\sqrt\hertz} and the noise floor is about \SI{0.1}{\femto\meter\per\sqrt\hertz}. The resulting RMS noise in the spectral range below \SI{1}{\kilo\hertz} is \SI{440}{\femto\meter} demonstrating the high performance of the vibration insulation of the instrument.  

\subsection{Magnetic excitations in single Fe atoms on Pt(111)}

Magnetic excitations of single Fe atoms on Pt(111) were first studied in 2009 using a STM operating at \SI{4.2}{\kelvin}~\cite{Balashov2009}. The second derivative of the tunneling current as function of voltage was measured on top of single atoms and interpreted as an antisymmetric peak-dip pair with the position of the peak corresponding to the excitation energy. Due to a limited energy resolution this interpretation led to a relatively high excitation energy in the meV range. In 2013 the system was revisited by A.~Khajetoorians et al.~\cite{Khajetoorians2013} using an STM at \SI{300}{\milli\kelvin} with much higher resolution. This study found significantly lower excitation energies for the two inequivalent adsorption sites on that surface.

\begin{figure}
\includegraphics[width=0.8\columnwidth]{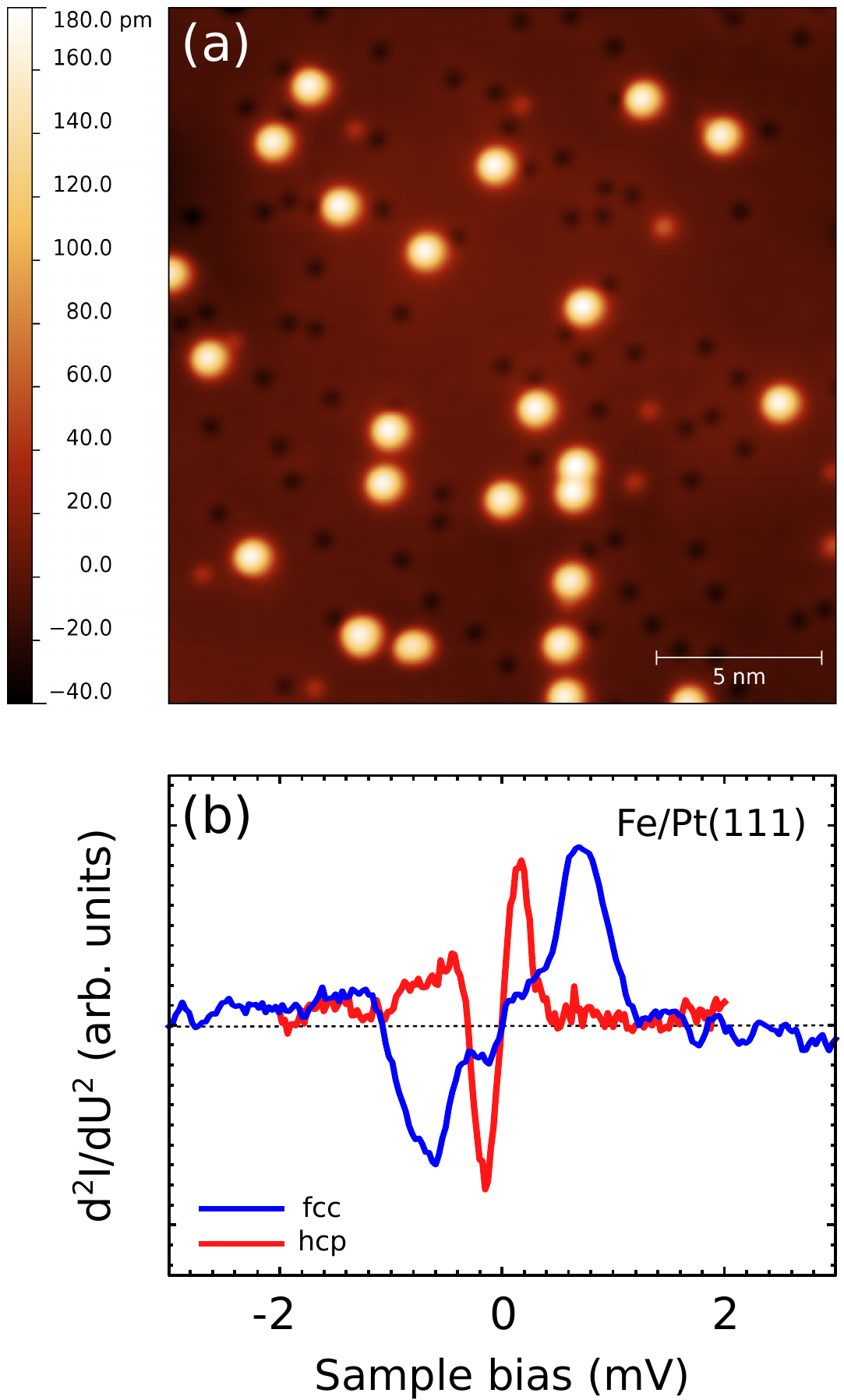}
\caption{(a) STM topography of single iron atoms on Pt(111). (b) $d^2I/dU^2$ signal showing inelastic excitations in Fe atoms on Pt(111) on two different adsorption sites measured at \SI{25}{\milli\kelvin} with \SI{0.5}{\milli\volt} RMS modulation voltage.}\label{fig_FePt1}
\end{figure}

\begin{figure}
\includegraphics[width=\columnwidth]{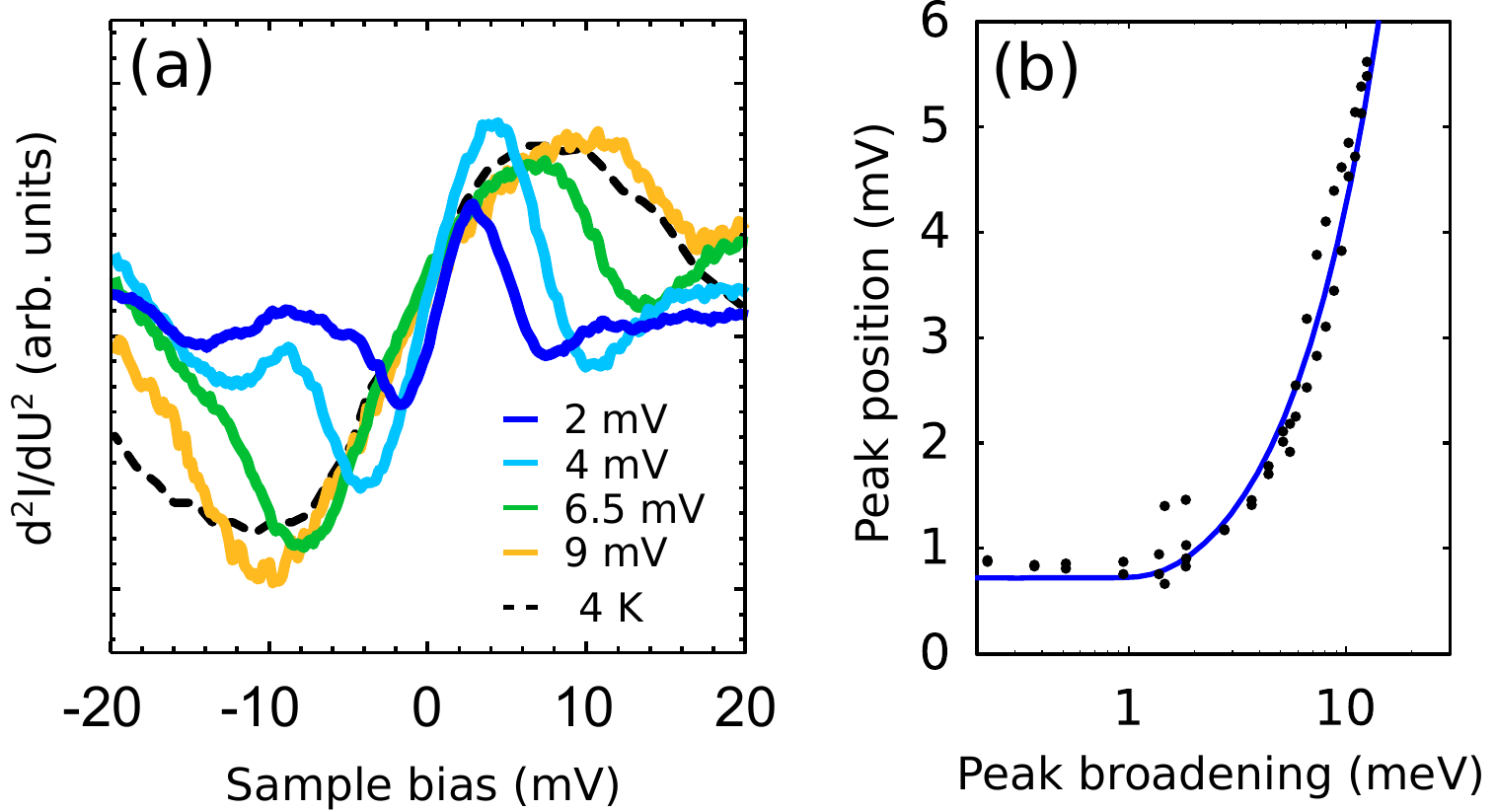}
\caption{The effect of energy resolution on inelastic exciation spectra. (a) With increasing modulation the inelastic peaks that were clearly resolved merge and the maximum of the curve shift to higher energies. At \SI{9}{\milli\volt} modulation the spectra become quite similar to the ones reported in ref.~\cite{Balashov2009} (black dashed line). (d) Experimental peak position as function of experimental broadening (black dots) closely follows the theoretical prediction (blue curve)~\cite{Balashov2014}.}\label{fig_FePt2}
\end{figure}

We have repeated the experiment in the new setup, depositing Fe atoms on clean Pt(111) below \SI{10}{\kelvin}, and measured $d^2I/dV^2$ spectra using a lock-in amplifier with \SI{750}{\micro\volt} modulation at \SI{14.3}{\kilo\hertz}.
The obtained data is presented in Fig.~\ref{fig_FePt1}. We have reproduced the results of Khajetoorians et al., observing two different sets of spectra, corresponding to different adsorption sites of Fe atoms. The position and width of the inelastic features correspond well to the values measured by Khajetoorians et al., and the spectra behave in the same way in magnetic field (not shown).

While is was argued that the observed excitation energies should not depend on the instrumental resolution \cite{Khajetoorians2013}, a theoretical analysis of the problem indicated the opposite \cite{Balashov2014}. Thus, we experimentally test the theoretical prediction of inelastic peaks in $d^2I/dV^2$ moving to higher energies with decreasing experimental resolution by increasing the modulation voltage (see Fig.~\ref{fig_FePt2}a). 
We have reproduced our results from 2009 by using a high modulation voltage of \SI{9}{\milli\volt}, mimicking higher temperature and noise of the old \SI{4}{\kelvin} setup (see Fig.~\ref{fig_FePt2}a and Ref.~\onlinecite{Balashov2014}). Moreover, the observed peak positions (black dots) closely follow that of the theoretical prediction (blue line) as function of peak broadening, i.e. instrumental resolution. These results indicate that inelastic peak-dip pairs can only give a reliable value for the excitation energy if the two features are clearly separated at zero bias. In the current study this is the case for Fe atoms on \textit{fcc} adsorption sites, and thus this excitation energy of $\SI{\approx700}{\milli\electronvolt}$ is reliable. The peak and dip are, however, not well separated in the spectra for Fe atoms on \textit{hcp} adsorption sites, and thus their position only indicates the upper limit for the excitation energy of \SI{\approx100}{\micro\electronvolt}. Although this value lies within our energy resolution capabilities, we were unable to separate the peak and dip reliably, which suggests that the intrinsic width of this excitation, i.e. its lifetime broadening,  is larger than the excitation energy and lies between 100 and \SI{200}{\micro\electronvolt}.

\section{Conclusion}

In conclusion, a compact UHV compatible STM operating at temperatures down to \SI{25}{\milli\kelvin} has been designed and tested. It allows a fast sample turnaround time below \SI{2}{\hour} and the application of magnetic fields up to \SI{7.5}{\tesla}. Sample transfer can be done under vision of the operator and single atoms can be deposited on cold surfaces. The instrument combines the easy use, small size, and low consumption of cryogenic liquids of conventional cryogenic STMs, with high magnetic fields and mK temperatures. This allows to study systems with the same throughput as with conventional STMs and renders pre-studies with other setups unnecessary. 

The instrument has been used to measure magnetic excitation spectra of 3d transition metal ions in magnetic molecules in magnetic fields both in the tunneling and the contact regime demonstrating a high energy resolution and high mechanical stability \cite{Chen2018}. It has further been used to measure the inelastic coupling between electrons and bosonic degrees of freedom in unconventional superconductors \cite{Jandke2018}.

\section{Acknowledgements}

The authors acknowledge funding by the German Research Foundation (DFG) under the grant INST 121384/30-1 FUGG.

\bibliographystyle{apsrev4-1}
\bibliography{dstmbib}
\end{document}